# Vehicle Powertrain Connected Route Optimization for Conventional, Hybrid and Plug-in Electric Vehicles

Zhiqian Qiao[1,2] and Orkun Karabasoglu[1,3*]

[1]Sun Yat-sen University – Carnegie Mellon University
Joint Institute of Engineering, Guangzhou, China
[2] Electrical and Computer Engineering Department
[3]Carnegie Mellon Scott Institute for Energy Innovation
Carnegie Mellon University, Pittsburgh, PA, US

* Author to whom correspondence should be addressed; E-Mail: karabasoglu@cmu.edu

***Abstract***

Most navigation systems use data from satellites to provide drivers with the shortest-distance, shortest-time or highway-preferred paths. However, when the routing decisions are made for advanced vehicles, there are other factors affecting cost, such as vehicle powertrain type, battery state of charge and the change of component efficiencies under traffic conditions, which are not considered by traditional routing systems. The impact of the trade-off between distance and traffic on the cost of the trip might change with the type of vehicle technology and component dynamics. As a result, the least-cost paths might be different from the shortest-distance or shortest-time paths. In this work, a novel routing strategy has been proposed where the decision-making process benefits from the aforementioned information to result in a least-cost path for drivers. We integrate vehicle powertrain dynamics into route optimization and call this strategy as “Vehicle Powertrain Connected Route Optimization (VPCRO)”. In order to show the cost benefits of VPCRO, the Dijkstra’s algorithm was employed to solve the least-cost problem on a traffic network model based on Shanghai, China with an average traffic information collected from Google Maps. The performance of the proposed routing strategy was compared to the shortest-distance and shortest time routing strategies. It was found that the optimal paths might change significantly for all types of vehicle powertrains. About 81% and 58% of trips were replaced by different optimal paths with the proposed VPCRO strategy when the vehicle type was Conventional Vehicle (CV) and Electrified Vehicle (EV), respectively. Changed routes had reduced travel costs on an average of 15% up to a maximum of 60% for CVs and on an average of 6% up to a maximum of 30% for EVs. Moreover, it was observed that 3% and 10% of trips had different optimal paths for a plug-in hybrid electric vehicle, when initial battery SOC changed from 90% to 60% and 40%, respectively. Paper shows that using sensory information from vehicle powertrain for route optimization plays an important role to minimize travel costs.

**Keywords:** Electric Vehicle, Vehicle Powertrain Connected Route Optimization, Plug-In Hybrid Electric Vehicle, Least-cost Routing, Navigation

## 1. INTRODUCTION

Navigation tools are heavily used by drivers in order to facilitate travel from one location to another. These tools depend on the solution of a route optimization problem. The problem formulation, the inputs and the solution strategies that are employed have the potential to change the optimal routing result as well as the cost of travel. Given the scale of transportation sector, even small improvements might lead to massive cost reductions. Thus, route optimization is a very important problem to be addressed.

Most traditional navigation systems in the market mainly provide shortest-distance, least time or highway-preferred routes. These methods do not address the cost of travel directly. Shortest path may be the least-cost path for conventional vehicles (CV), if the efficiency of the vehicle remains constant, which is not the case in reality. Literature shows that the efficiency of the vehicles are not constant and might change due to several factors such as powertrain architecture [1], component properties [2], traffic [3], driving patterns [4], terrain [5], control strategy [6], air conditioning being on/off [7] etc. This ultimately changes the cost of each segment, and a result optimal routing decision might change.

Furthermore, with the introduction of electrified vehicles such as hybrid, plug-in and battery electric vehicles, interaction of vehicle powertrain dynamics and traffic conditions will have a bigger impact on the cost of the segments of the transportation network. The conventional vehicles work with gasoline engines that suffer from low efficiencies under varying driving conditions [8]. In order to keep the gasoline engine under efficient operation range, it can be paired with an electric motor and a small battery pack to assist the engine. This type of vehicle powertrain is called Hybrid Electric Vehicle (HEV). The increased efficiency results from mainly: (1) keeping the engine on high efficiency operating curve, (2) making it possible to use smaller and higher efficient engines due to the tandem use of motor and engine, and (3) regenerative braking feature which uses the electric motor as a generator to restore some of the excess mechanical energy during braking back to the battery. When HEVs integrated with large rechargeable batteries, it is called Plug-in Hybrid Electric Vehicle (PHEV). PHEVs can travel partially or entirely on electricity, which is cheap compared to fuel. Battery Electric Vehicles (BEVs) do not have a gasoline engine, and can only travel on electricity using their motor and rechargeable large battery pack. For plug-in hybrid electrified vehicles, the cost of each road segment of a trip depends on both the fuel and electricity consumed to cover it.

Least-cost optimal paths might be different for different vehicle powertrains (CV, HEV, PHEV, and BEV) due to their varying efficiency in different traffic conditions. For example, traditionally a conventional vehicle would avoid heavy traffic since gasoline engines have low efficiencies in stop and go driving conditions. Given two routes, one a little longer with less traffic and the other one, shorter with more traffic; the least-cost route for a CV might be the longer path as a result of the trade-off between distance and traffic. However, a hybrid electric vehicle might perform better in heavier traffic conditions due to a high efficient engine and regenerative braking ability. As a result, a HEV could choose the shorter path with more traffic compared to a CV which would chose longer path with less traffic.

Similarly, initial battery state of charge in electrified vehicles might change the cost-minimum route. All Electric Range (AER) is defined as the distance that vehicle can travel before switching to consuming gasoline. Initial battery state of charge might change AER, the portion of the trip that can be traveled only on electricity. AER will have a big impact on the cost of the trip since electrical power is much cheaper compared to gasoline.

In this paper, we propose receiving information regarding vehicle dynamics from the Control Area Network (CAN) bus of the vehicle through On Board Diagnostics (OBD) port of the vehicle and integrating the varying powertrain component efficiencies into optimal routing algorithm. Then we compare the results of our proposed method to that of shortest path and least time algorithms for several vehicle powertrains types such as CV, HEV, and plug-in vehicles such as PHEVs and BEVs with different initial battery state of charge.

The proposed system consists of four parts. The first part is a sensory system that collects data about the state, technology and efficiency of the vehicle powertrain for given or anticipated traffic conditions (such as battery size, state of charge, vehicle powertrain type, efficiency, speed, acceleration, altitude). This can be an OBD (On Board Diagnostics) scanner, or any vehicle-in-built sensor. The second part is a data collection system such as a smart phone or navigation system which can collect data from the sensor as well as from the user and transfer it to the cloud computing. The third part is all the other data such as map of transportation network, GPS coordinates, unit price of energy, user preferences (starting location, final destination) and etc. After the results of optimal routes feeds back to the core device, the equipment displays the navigation in detail to users. The last part of the system is a cloud computing system where the vehicle connects to for route optimization. By using the integrated data and designed algorithm, the server searches the best route and feeds it back to the driver.

In the next session, we present the relevant literature and then detail our proposed framework and analysis.

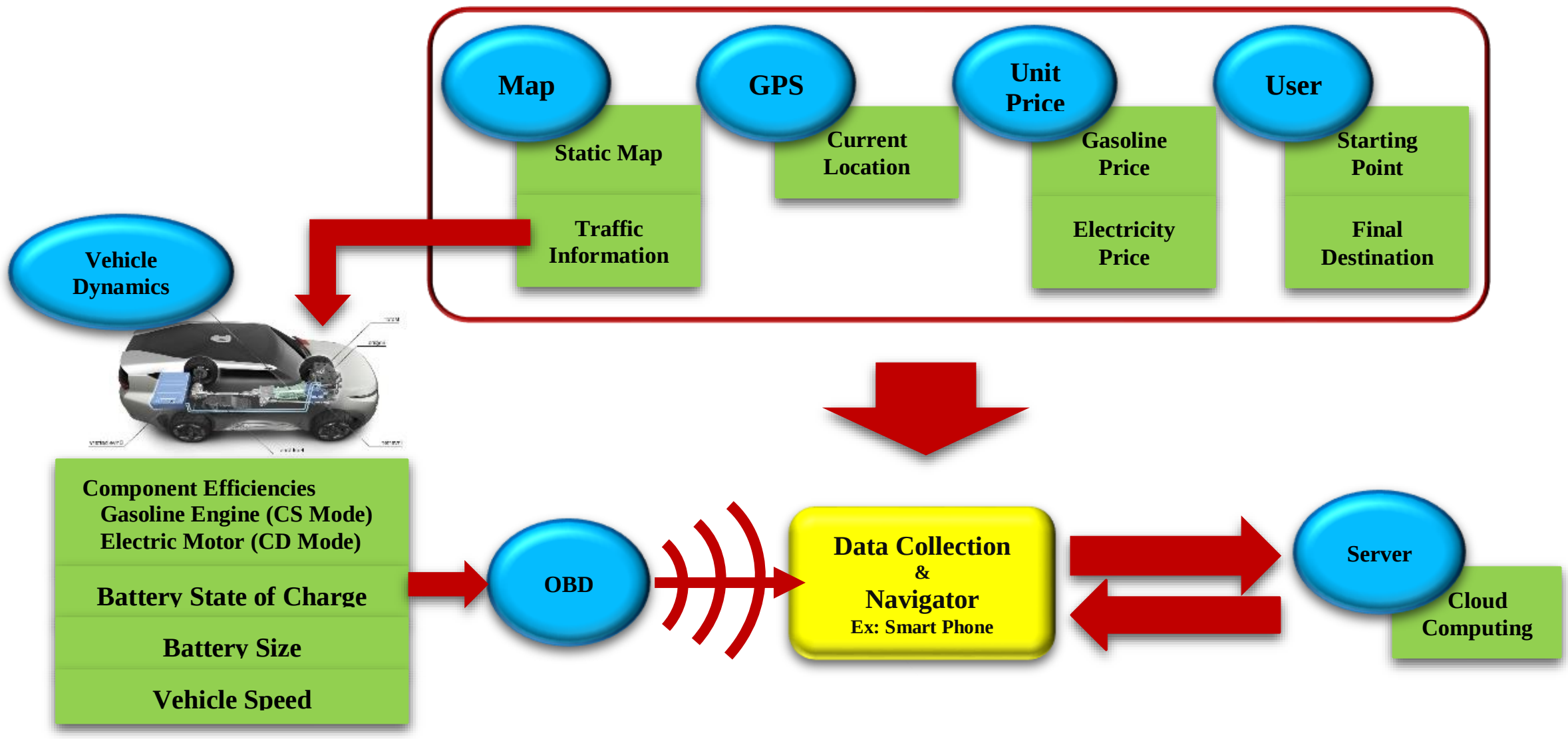

**Figure 1:** The Proposed Framework for Vehicle Powertrain Connected Route Optimization (VPCRO)

## 2. LITERATURE REVIEW

Literature on route optimization can be classified into: (1) studies that focus on route selection for conventional vehicles under traffic conditions, (2) studies that investigate different objective functions for conventional vehicles such as minimizing fuel consumption, travel time, and total cost; and (3) development of routing systems for electric vehicles. We review each category with detail in the following paragraphs.

There are works that focus on optimal routing through shortest path in conventional vehicles. Traditionally, route optimization for conventional vehicles has been done through solving shortest path (SP) problems. For the solution of SP, there have been different algorithms proposed such as Dijkstra's Algorithm [9], A*[10], Genetic Algorithms [11], Improved Bellman–Ford Successive Approximation Algorithm [12], Particle Swarm Optimization [13] and Column Generation techniques [14].

In addition to the optimal routing using the shortest path, researchers have considered other factors that affect the efficiency of the vehicle and change the cost of travel. These factors might be traffic, driving patterns, terrain, vehicle load, air conditioner load and etc.

Karabasoglu and Michalek [4] investigated the effect on driving patterns on life cycle cost and emissions of different categories of powertrains under various scenarios and simulated driving conditions. They found that driving patterns matter and have the potential to change the ranking of advanced powertrains for their benefits. Zhang et al. [5] integrated information of road terrain with the energy management systems of hybrid vehicles and showed that the incorporation of terrain review can help reduce energy consumption. Xiao et al. [15] investigated the impact of vehicle load on the classical capacitated vehicle routing problem and found that load is an important factor to consider. Tavares et al. [16] showed that vehicle weight and the inclination of roads affects the efficiency of the vehicle. Their routing strategy increased cost savings by 8% when compared to the three dimensional routes obtained by a method which did not consider terrain review. Kanoh and Hara [17] devised a dynamic route planning strategy to minimize travel time and length of the route, which performed better than the conventional multi-objective Genetic Algorithms. Ericsson et al. [18] calculated the shortest path on the basis of fuel economy and their strategy saved fuel on an average of 8.2%. Ahn et al. [19] investigated the impacts of route choice decisions on vehicle energy consumption and emission rates of conventional vehicles. They found that the route choice of faster highway is not always the best from an environmental and energy consumption perspective. Scora et al. [20] considered the effect of vehicle mass and terrain on fuel consumption and optimal routing. They state that neglecting of road grade reduces fuel consumption estimation as much as 12.7% for a selected route. Saboohi and Farzoneh [21] studied the influence of the parameters like vehicle speed and gear ratio on optimal routing. Taking the optimal route proposed by this system could result in fuel savings up to 1.5 L in 100 km during congested traffic conditions.

In the recent years, with the introduction of electrified vehicles, researchers have been working on extending the conventional routing strategies in order to adapt to the changing nature of vehicle fleet. Artmeier et al. [22] proposed extensions to general shortest-path algorithms that addressed the problem of energy-based-optimal routing in the presence of rechargeable batteries with the help of a weighted graph where the weights represented the energy consumption in the network. He et al. [23] developed a mathematical model which takes into account the relations among the choices for the route and the parameters like travel time, availability of charging stations, and cost of charging. Baum et al [24] minimized travel time by using energy optimal route planning in electric vehicles utilizing variable cost functions influenced by changing weather conditions. Sweda et al. [25] proposed a model, which considered finding a path for an EV which offered minimum cost when the vehicle recharges along the way. Recharging decisions were considered in this paper when planning EV routes since they have significant impacts on the total travel time and longevity of the batteries, which in turn will incur more costs during replacement. Schneider et al [26] devised a vehicle routing problem with recharging stations that minimized the total driving distance and outperformed the performances of the conventional routing strategies devised by Erdogan and Miller-Hooks [27]. Jurik et al. [32] presented an Energy Optimal Real Time Navigation System which considers optimal energy management and optimal routing simultaneously for supercapacitor-battery electric vehicles. A robust optimization based framework incorporating uncertainty due to different factors like traffic conditions, weather, modeling complexities, was developed in [33] for solving the vehicle routing problem. They utilized a robust energy model to ensure that the routing process terminates with sufficient energy in the battery.

Vehicle routing process might benefit from using input from vehicle powertrain, component states and their interaction with the traffic conditions. In this study, we propose a framework where we integrate vehicle dynamics into routing strategy and call it "Vehicle Powertrain Connected Route Optimization (VPCRO)". Then we analyze the impact of our proposed strategy on cost reductions and change of routes for different vehicle powertrains such as CV, HEV, PHEV and BEV.

## 3. METHODOLOGY

In this section, we introduce the methodology for our proposed VPCRO strategy which integrates the vehicle powertrain dynamics, component efficiencies, control modes, energy storage system specifications and their initial states into routing optimization. These factors affect the efficiencies of powertrain components which ultimately changes the cost of travel on each segment of the trip. Similarly, vehicle type, battery size, initial battery SOC along with traffic flow determines what percentage of the trip will efficiently be covered on electricity or gasoline. The proposed routing strategy in this paper, illustrated in Figure 1, takes into account all the aforementioned factors and provides vehicle powertrain and initial system state specific least-cost optimal path. Then we compare the results of our strategy with traditional shortest-distance and shortest-time methods for conventional vehicles and electrified vehicles. In order to simulate these routing strategies, the first step is to construct a network model with traffic information. Then we build vehicle models, formulate and solve the optimization problem and analyze the results.

### 3.1. Construction of transportation network

To construct the transportation network, node and segment data were collected from the map of Shanghai, China and average traffic condition is mapped using the historical data from Google Maps during 8:00 a.m. to 8:30 a.m. on a Monday. Figure 2 illustrates the map of Shanghai from Google Maps and the constructed traffic network for simulation purposes. The network consists of 352 nodes and 615 segments representing intersections and roads respectively, where a total of 61776 Origin-Destination (O-D) pairs can occur. Traffic conditions in the network have been categorized into low, average and heavy traffic. The red, yellow and green segments in Figure 2 represent heavy, average and low traffic respectively. Traffic conditions collected from Google Maps can be represented by certain driving cycles since traffic flow is one of the most important factors that dictate the speed of the vehicle over time. Here, it is assumed that the traffic conditions of each segment can be approximated by the following driving cycles: For low traffic conditions, Highway Fuel Economy Test (HWFET) driving cycle has been employed while for the average traffic conditions, Urban Dynamometer Driving Schedule (UDDS) driving cycle has been chosen. For heavy traffic, the New York City (NYC) driving cycle has been used since it reflects heavy traffic with frequent stops.

The transportation network is represented by a directed graph **D**=**{X, S}**, where **X** represents the nodes and **S** represents the segments which are directed and bi-directional. The segment $s_{ij} \in \mathbf{S}$ connects node $n_i$ to $n_j$. The set **X** consists of 352 nodes in the traffic network model and each node is assigned an index number *i* and associated with the corresponding 2-D coordinates collected from Google Maps. The 2-D coordinates are shown as longitude $n_i^{\text{lon}}$ and latitude $n_i^{\text{lat}}$ in unit meters for the node $n_i$.

The set **S** contains 615 segments $s_{ij}$ of the network. Each segment is associated with three kinds of weight information $w_{s_{ij}}^d$, $w_{s_{ij}}^t$, $w_{s_{ij}}^c$ which corresponds to distance of segment $d_{s_{ij}}$, travel time of segment $t_{s_{ij}}$ and cost of the segment $c_{s_{ij}}$ due to the total fuel and energy consumption on the segment. These weight factors are utilized in vehicle route optimization, which is covered in detail in Section 3.3.

$$w_{s_{ij}}^d = d_{s_{ij}}, w_{s_{ij}}^t = t_{s_{ij}}, w_{s_{ij}}^c = c_{s_{ij}} \quad (1)$$

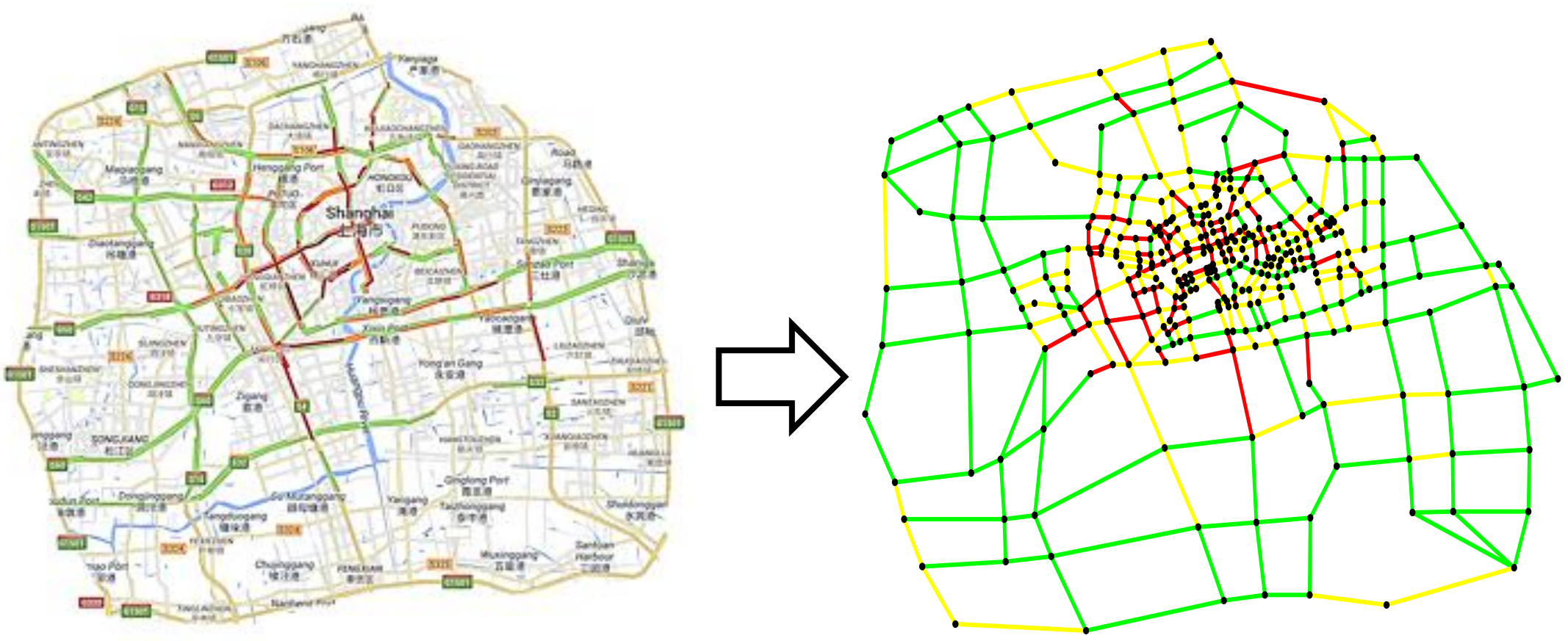

**Figure 2:** Traffic Network Model of Shanghai

The distance, $d_{s_{ij}}$, between any two nodes $n_i$ and $n_j$ is defined in the Equation 2. $C_d$ = 62.137 mile/m is the scaling factor used to convert meters to miles.

$$\text{If } n_i \text{ and } n_j \text{ are connected } \; d_{s_{ij}} = C_d \sqrt{(n_i^{\text{lon}} - n_j^{\text{lon}})^2 - (n_i^{\text{lat}} - n_j^{\text{lat}})^2} \qquad (2)$$

$$\text{Else } \; d_{s_{ij}} = \infty$$

### 3.2. Advanced Vehicle Powertrain Modeling

In this paper, six types of vehicle powertrains (CV, HEV, PHEV20, PHEV40, PHEV60, and BEV100) are simulated for three kinds of routing strategies: shortest distance, least time and the proposed VPCRO least cost strategy. Since different powertrain components such as gasoline engine, electric motor and battery might have different efficiencies under various scenarios such as traffic conditions, the travel cost of each segment may change for different powertrains. We use Powertrain System Analysis Toolkit (PSAT) to design and simulate the vehicles. We adopted vehicle models from [4]. Table 1 lists the component sizes of the vehicle powertrains used in this paper. The CV and HEV vehicle models that are used in the paper are based on Toyota Carolla and Toyota Prius, respectively. The PHEV models are based on the MY13 PHEV configuration with increased battery size and the BEV model is constructed with a modified mid-size electric powertrain in PSAT. F/R weight ratio is 06/04, drag coefficient is 0.26, frontal area is 2.25 m2. HEV initial SOC and target SOC are set to 60%. For PHEVs and BEVs, in CD mode the initial SOC is set to 90% and target SOC is set to 30%, and for CS mode the initial and target SOC are set to 30%. More details of the vehicle assumptions and efficiency maps can be found in [4].

**Table 1:** Vehicle Configurations [4]

| Vehicle Type | Engine(kW) | Motor(kW) | Battery(kWh) | Mass(kg) |
|---|---|---|---|---|
| CV | 110 | - | - | 1371 |
| HEV | 73 | 60 | 1.3 | 1424 |
| PHEV20 | 73 | 78 | 9.9 | 1569 |
| PHEV40 | 73 | 88 | 19.9 | 1793 |
| PHEV60 | 73 | 98 | 30.2 | 2027 |
| BEV100 | - | 120 | 54.0 | 2265 |

For each vehicle, the most significant variable for routing process is the different efficiencies of the powertrain when driving through various road conditions. For CV and HEV, the efficiencies of vehicles vary with traffic conditions. However, for PHEV, the distances of trips as well as traffic conditions impact the energy consumption of vehicles. Plug-in vehicles operate on charge depleting (CD) mode at the beginning of a trip, and when the battery target SOC is reached, vehicles automatically switch to using gasoline, which is called Charge Sustaining (CS) mode. BEV, which is only powered by the battery, will eventually use all the available energy and run out of electricity energy and will need a recharge or battery swap. The most energy demanding possible trip in the simulated transportation network does not consume all the available energy of battery pack in BEV100, so we assume that it operates only under CD mode during the whole trip.

To evaluate if the vehicles have entered into CS mode and started consuming gasoline, remaining energy and target SOC is used. The initial energy of battery differs for PHEVs with different battery sizes. Ignoring all the other factors which impact the remaining electric power, Figure 3 shows the relation between available energy and driving distances for PHEVs. All Electric Range (AER) refers to the distance an electrified vehicle can cover by only using electricity before switching to another energy source when available. For example, PHEV20 refers to a PHEV design that can only cover 20 miles on electricity. AER might change depending on the driving patterns up to 40% [4] due to the interaction of vehicle powertrain components with the traffic and driving conditions.

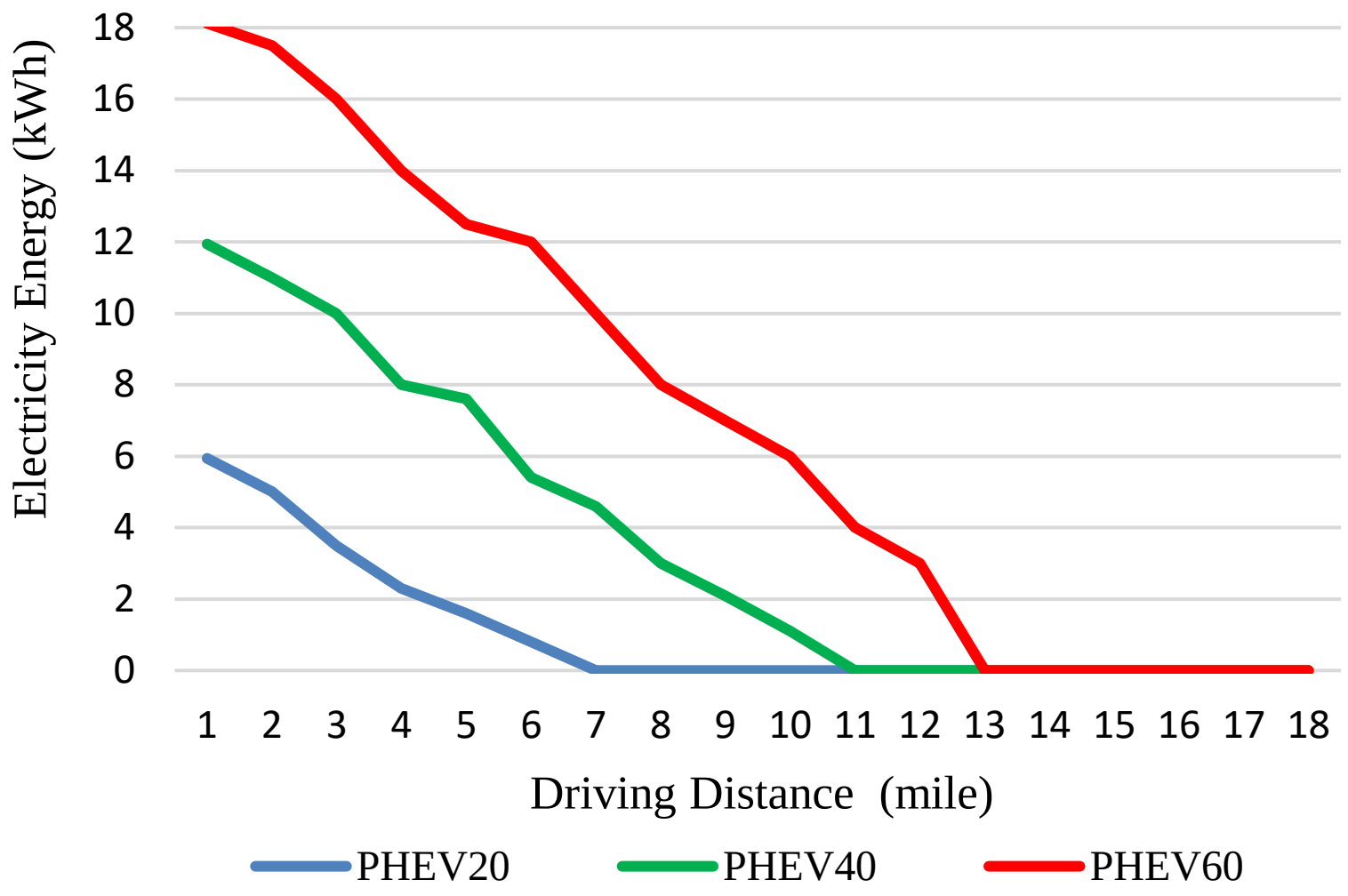

**Figure 3:** Relation between Remaining Electric Energy and Driving Distance

By modeling the vehicle configurations of Table 1 in PSAT, we can simulate powertrains and calculate their conversion factors under various driving cycles. Table 2 shows the conversion factor μ of vehicles under CD and CS modes, and the initial electricity energy $E_{ini}$ of different powertrains under each of the three driving cycles proposed [4].

**Table 2:** Conversion factor of different vehicle powertrains [4]

| Vehicle Type | Symbol | Unit | HWFET | UDDS | NYC |
|---|---|---|---|---|---|
| CV | $\mu_{CS}$ | mi/gal | 52.8 | 32.1 | 16.4 |
| HEV | $\mu_{CS}$ | mi/gal | 59.7 | 69.5 | 48.0 |
| PHEV20 | $\mu_{CD}$ | mi/kWh | 5.7 | 6.2 | 4.2 |
| | $\mu_{CS}$ | mi/gal | 58.6 | 69.4 | 45.7 |
| PHEV40 | $\mu_{CD}$ | mi/kWh | 5.7 | 6.0 | 4.1 |
| | $\mu_{CS}$ | mi/gal | 58.2 | 68.0 | 43.1 |
| PHEV60 | $\mu_{CD}$ | mi/kWh | 5.6 | 5.7 | 3.8 |
| | $\mu_{CS}$ | mi/gal | 57.8 | 65.8 | 40.3 |
| BEV100 | $\mu_{CD}$ | mi/kWh | 4.8 | 5.2 | 3.1 |

**Table 3:** Initial available energy of different vehicle powertrains [4]

| Vehicle Type | $E_{ini}$ (kWh) |
|---|---|
| PHEV20 | 5.94 |
| PHEV40 | 11.94 |
| PHEV60 | 18.12 |

Another key factor which influences the efficiencies of plug-in vehicles is the initial battery state of charge (SOC). Initial battery SOC has a direct impact on the available energy $E_{ini}$ and indirect effect on the portion of the trip that has been travelled under CD mode and CS mode.

If a driver starts their trip with a battery that does not contain enough energy, they will soon run out of energy and vehicle will switch to CS mode earlier than expected. In this paper the target SOC for the PHEV and the HEV has been considered to be 30%. Table 4 presents the unit price of gasoline and electricity.

**Table 4:** Unit Cost of Energy [4]

| $p_{ele}$ (\$/kWh) | $p_{gas}$ (\$/gallon) |
|---|---|
| 0.114 | 2.75 |

### 3.3. Route Optimization

Least cost route optimization for advanced vehicles necessitates the information about component efficiencies such as engine and motor, energy prices, traffic conditions, powertrain types, control modes and initial battery state of charge. Ultimately these factors change the vehicle conversion factors. We can define the route optimization problem with the following objective functions: the shortest distance, the shortest time and the least cost as shown in the equations 3, 4 and 5, respectively.

$$\textbf{\textit{Shortest Distance}} \quad \underset{s_{ij}}{minimize}\left(\sum_{(i,j)\in\mathbf{R}} w_{s_{ij}}^{d}\right) \tag{3}$$

$$\textbf{\textit{Shortest Time}} \quad \underset{s_{ij}}{minimize}\left(\sum_{(i,j)\in\mathbf{R}} w_{s_{ij}}^{t}\right) \tag{4}$$

$$\textbf{\textit{Least Travel Cost}} \quad \underset{s_{ij}}{minimize}\left(\sum_{(i,j)\in\mathbf{R}} w_{s_{ij}}^{c}\right) \tag{5}$$

where **R** is the set of possible paths from the initial node to the goal node; and *i* and *j* are the indices of the nodes in the path. We define the weight of each segment between any two nodes $n_i$ and $n_j$ as $w_{s_{ij}}^{d}$, $w_{s_{ij}}^{t}$ and $w_{s_{ij}}^{c}$ for distance, time and cost optimization, respectively.

In order to solve the route optimization of the 61776 possible trips in our simulation network, we employ Dijkstra's algorithm [9]. The weight between each node of network is non-negative due to the fact that all the distances, time and travel cost of each segment being positive. Dijkstra's algorithm is suitable for the solution of single-source shortest-path problems for arbitrary directed graphs with unbounded non-negative weights. In order to compare our routing strategy against the shortest-distance and shortest-time strategies, we use Dijkstra's algorithm for all three objectives separately.

In the Dijkstra's algorithm, the input is the digraph **D={X, S}**, where **X** represents the nodes and **S** represents the segments which are directed and bi-directional. Each segment has a designated weight $w_{s_{ij}}^{k}$, and the subscript $k$ represents the objective to be minimized over the segments $s_{ij}$ on the route from origin to destination. Initially, the implementation of the algorithm depends on the labelling of the nodes with a potential $W^{k}(j)$ in the set **X**. The value of the potential $W^{k}(j)$ will be finite for any two general nodes $n_A$ and $n_B$ ($n_A$, $n_B$ ∈**X**) which are linked with a segment in the graph **D**. For $n_{\mathrm{m}}$ representing the elements of the set **N** ($n_{\mathrm{m}}$ ∈ **N**) consisting of the starting nodes, the potential will be zero. For all other nodes having no direct linked segments, $W^{k}(j)$ =∞. Let **V** represent the set of elements $n_v$ consisting of the destination nodes ($n_v \cong n_m$), **P** present the visited nodes and **Q** represent the unvisited nodes. **P** initially includes the starting node $n_m$. From the initial node $n_m$, all the unvisited neighbors are considered and their respective potential $W^{k}(j)$ are updated as the tentative weights ($w_{s_{ij}}^{k}$) of the interlinked segment. This process indicates a visit to a new node or a re-visit to a previously processed node via the shortest path. In the upcoming steps, the update rule is only valid if the current values of the potential are smaller than the previous ones. After the following updates are completed, the neighboring node with smallest potential, is allocated to the set **P** after being removed from **Q**. In the next iteration, the allocated node is then denoted as the starting node and the process continues till **P** keeps all the nodes bearing finite values of the potential $W^{k}(j)$ and **Q** keeps the rest. The iteration process terminates when the set **Q** is empty and all the nodes in **D** are labelled with their respective potentials.

So for each starting point $n_m$ ∈ **X**, we command this simulation process $R_m$ for three routing strategies separately. For a vehicle routing problem the objective can be shortest-distance, shortest-time or least-travel-cost between the initial node $n_m$ and the arbitrary destination node $n_j$. In this paper we denote the potentials by $W_f^{d}(j)$, $W_f^{t}(j)$ and $W_f^{c}(j)$ where $n_j$ ∈ **V.**

Then, the process $R_m$ of Dijkstra's algorithm for a starting node $n_m$ is:

$$
\begin{aligned}
W_0^k(i) &= w_{s_{mi}}^k \\
W_1^k(j) &= \min_{s_{ij} \in S} \{w_{s_{ji}}^k + W_0^k(i)\} \\
W_2^k(j) &= \min_{s_{ij} \in S} \{w_{s_{ji}}^k + W_1^k(i)\} \\
&. \\
&. \\
W_f^k(j) &= \min_{s_{ij} \in S} \{w_{s_{ji}}^k + W_{n-1}^k(i)\}
\end{aligned}
\tag{6}
$$

The process $R_m$ will end when $W_n^k(j)$ stops changing:

$$W_f^k(j) = W_{f-1}^k(j) \tag{7}$$

$W_f^k(j)$ represents the minimum cost of going from $n_m$ to $n_j$ where the superscript $k$ can be $d$, $t$ or $c$ to represent shortest-distance, shortest-time and lowest-cost strategies, respectively, the subscript $f$ represents the iteration count, with $f$ being final, and $j$ represents the next node $n_j$. The initial value of m is 1. After the process $R_m$ ends, the process $R_{m+1}$ will begin for another starting node in the network. Once all the starting nodes are covered, a lookup table consisting of three different cost values for distance, time and travel cost for all possible O-D (Origin Destination) combinations is generated. For calculating the time required to cover a segment, a navigation system generally applies the average speed of the segment to calculate how long it will take to go through that segment. In this paper, for comparability of three routing strategies, we apply average traffic information into time calculation when planning the routes. Average speeds $v_{s_{ij}}$ of different driving conditions approximated by UDDS, HWFET and NYC cycles have been tabulated in Table 5.

**Table 5:** Average Speed of Driving Cycles [4]

| | Unit | UDDS | HWFET | NYC |
|---|---|---|---|---|
| $v_{sk}$ | mph | 19.58 | 48.28 | 7.05 |

The time required to travel a segment $t_{s_{ij}}$ is calculated using the following formula:

$$t_{s_{ij}} = \frac{d_{s_{ij}}}{v_{s_{ij}}} \tag{8}$$

Where $d_{s_{ij}}$ is the segment length and $v_{s_{ij}}$ is the average speed data from Table 5.

We give the weights $w_{s_{ij}}^{k}$ for an arbitrary segment $s_{ij}$ for shortest distance, shortest time and least-cost routing below:

| | | | |
|---|---|---|---|
| ***Shortest Distance*** | | $w_{s_{ij}}^{d} = C_d\sqrt{(n_i^{\text{lon}} - n_j^{\text{lon}})^2 - (n_i^{\text{lat}} - n_j^{\text{lat}})^2}$ | (9) |
| ***Shortest Time*** | | $w_{s_{ij}}^{t} = \frac{d_{s_{ij}}}{v_{s_{ij}}}$ | (10) |
| ***Least Cost*** | **CV** | $w_{s_{ij}}^{c} = \text{p}_{\text{gas}}\frac{d_{s_{ij}}}{\mu_{CS}}$ | (11) |
| | **HEV** | $w_{s_{ij}}^{c} = \text{p}_{\text{gas}}\frac{d_{s_{ij}}}{\mu_{CS}}$ | (12) |
| | **PHEV** | $w_{s_{ij}}^{c} = \begin{cases} p_{gas}\frac{d_{s_{ij}}}{\mu_{CS}}, \text{ If } \quad \text{E}_{\text{rem}} \leq 0 \\ p_{ele}\frac{d_{s_{ij}}}{\mu_{CD}}, \text{ Else If } \quad E_{rem} \geq \frac{d_{s_{ij}}}{\mu_{CD}} \\ \text{p}_{\text{ele}}\, E_{rem} + \text{p}_{\text{gas}}\frac{d_{s_{ij}} - \mu_{\text{CD}}E_{rem}}{\mu_{\text{CS}}}, \text{ otherwise} \end{cases}$ | (13) |
| | **BEV** | $w_{s_{ij}}^{c} = \text{p}_{\text{ele}}\frac{d_{s_{ij}}}{\mu_{CD}}$ | (14) |

The conversion factors μ are taken from Table 2 for charge sustaining (CS) and charge depleting (CD) modes of different powertrains. The conversion factor under CD mode and CS mode have been represented by $\mu_{CD}$ and $\mu_{CS}$, respectively. Equations 11, 12, 13 and 14 give the cost of each segment $c_{s_{ij}}$ for CV, HEV, PHEV and BEV respectively. $p_{gas}$ and $p_{ele}$ denotes the gasoline and electricity cost, respectively and have been taken from Table 4. The remaining battery energy of plug in electric vehicles is denoted as $E_{rem}$ and can be represented below in equation 15.

$$E_{rem} = E_{ini} - \frac{W_f^c(j)}{\text{p}_{\text{ele}}} \tag{15}$$

# 4. RESULTS AND ANALYSIS

In this section, we compare the results of the VPCRO algorithm, which is proposed in this paper, to that of traditional shortest path routing and least-time routing algorithms. In Section 4.1, the influence of vehicle powertrain type on the least-cost optimal routing is discussed. Then in Section 4.2, the impact of initial battery state of charge on optimal routing and travel cost is demonstrated. Section 4.3 discusses the impact of VPCRO on travel time for different vehicle powertrains.

## 4.1. Impact of vehicle powertrain type on optimal routing

Vehicle type might affect the preference of routes to minimize travel costs. Different powertrains that consist of different traction components might have significantly different efficiencies in certain traffic situations compared to each other.

### 4.1.1 Percentage of changed routes and maximum cost savings under VPCRO

Figure 4 shows the percentage of changed routes, average and maximum cost reductions under VPCRO compared to the shortest distance and shortest time strategies for different powertrain types. About 80% and 60% of the trips have new optimal paths using the proposed optimization algorithm compared to shortest-distance algorithm for conventional and electrified vehicles, respectively. This implies that the majority of trips have been benefited from VPCRO strategy. However, compared to shortest-time strategy, about 45% and 85% of the trips have new optimal paths for CV and electrified vehicles, respectively. Furthermore, the proposed strategy might reduce costs up to 60% and 30% for CV and electrified vehicles (HEV and PHEV20), respectively when compared to shortest-distance strategy. The upper bound of cost savings for VPCRO compared to shortest-time are 45% and 73% for CV and electrified vehicles, respectively. The figure also shows the average cost reduction for the changed routes achieved for different vehicle types using the proposed strategy. Compared to the shortest distance routing strategy, VPCRO achieves 15% and 6% average cost reductions on the changed routes for CVs and BEVs, respectively

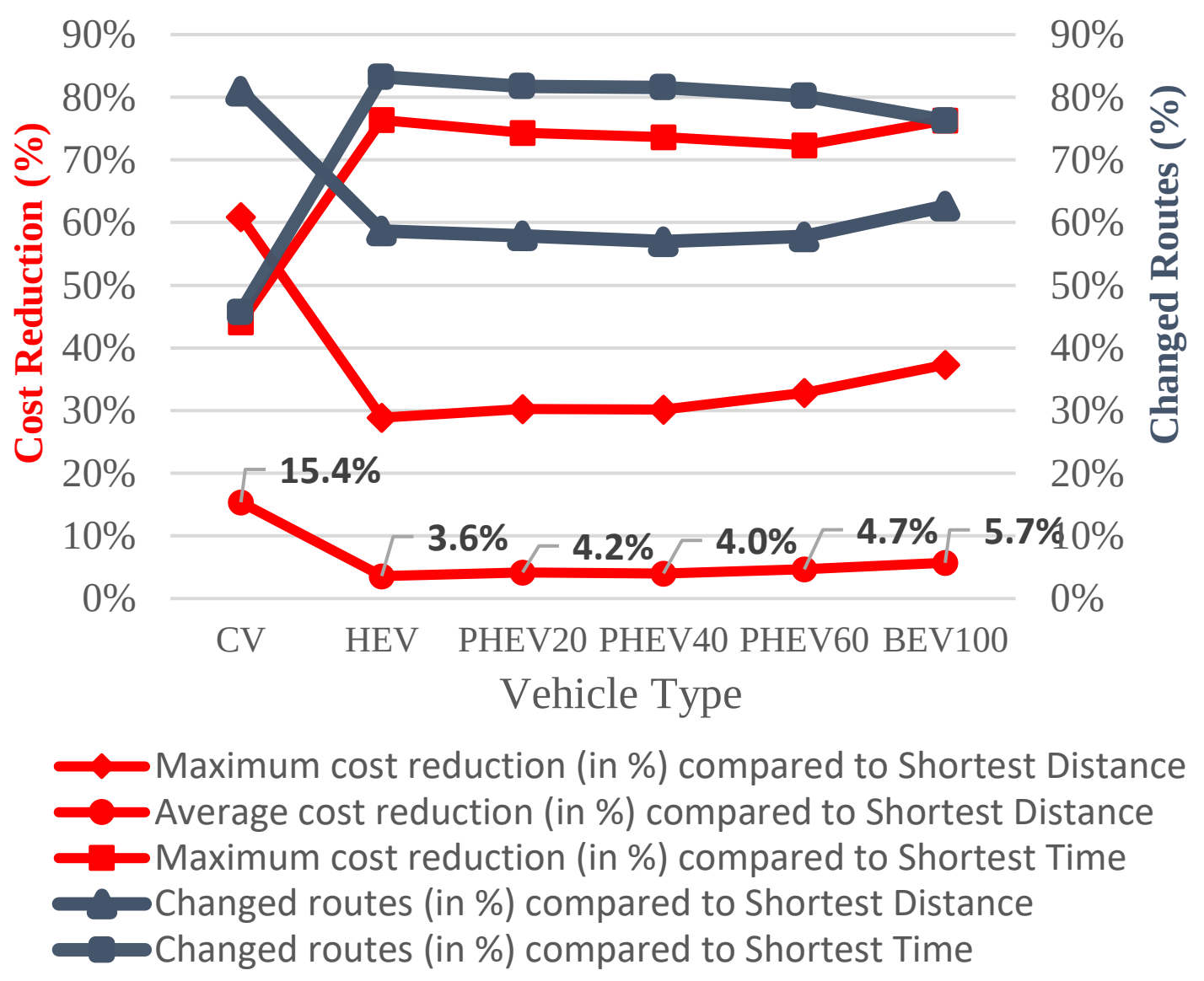

**Figure 4:** Cost reductions and percentage of changed routes under VPCRO compared to shortest distance and shortest time strategies for different powertrain types.

### 4.1.2 Distribution of cost savings compared to other routing strategies

Figure 5 shows the distribution of cost savings via VPCRO compared to Shortest-Distance and Shortest-Time for different vehicle types. From the histogram (Fig 5a), 80% and 60% of trips benefit from the VPCRO at different cost savings for conventional and electrified vehicles, respectively. Compared to least time routing (Fig 5b), VPCRO benefits 55% and 80% of the trips at different cost savings for conventional and electrified vehicles, respectively. While the powertrain efficiency increases (from CV to electrified vehicle), we find that the cost saving benefits of VPCRO is higher compared to shortest-time and lower compared to the shortest-distance methods.

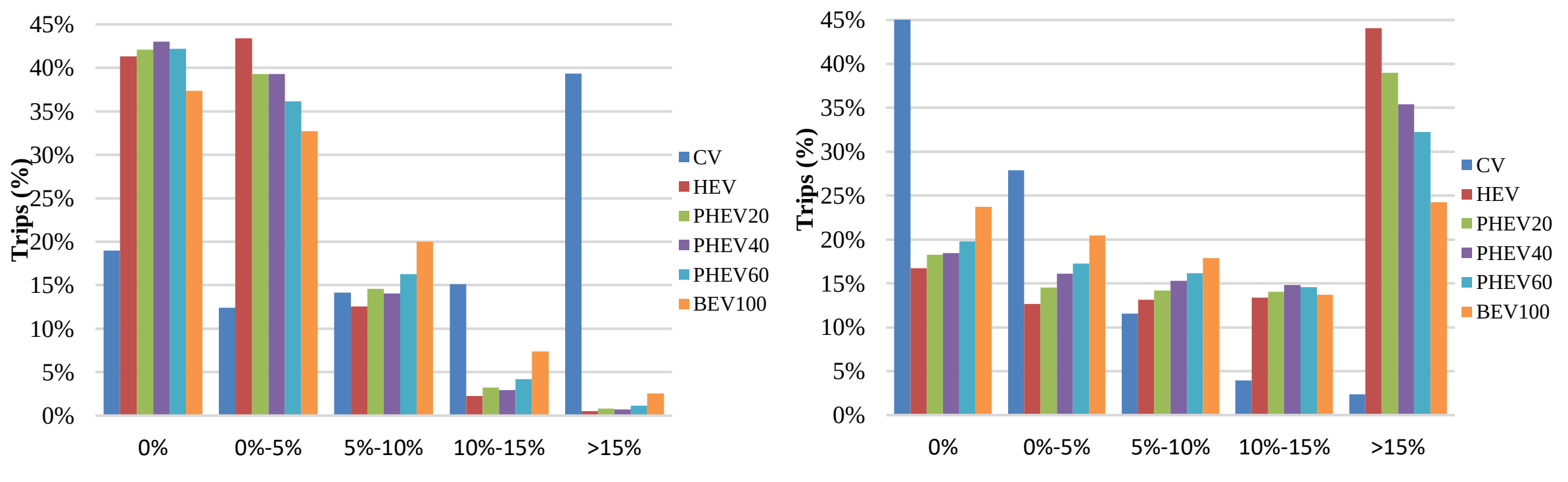

(a) (b)

**Figure 5**.(a) Distribution of cost savings of VPCRO compared to Shortest-Distance.
(b) Distribution of cost savings of VPCRO compared to Least-Time for different vehicle types.

### 4.1.3. Overview of changed routes

Figure 6 shows that the least cost path might be different for different vehicle powertrains. From a starting node to a destination node in the Figure 6, we demonstrate optimal routes: the blue (1) path represents the route found by shortest-path strategy, while the red (2) and green (3) paths represent the least-cost paths for CV and HEV found by VPCRO, respectively.

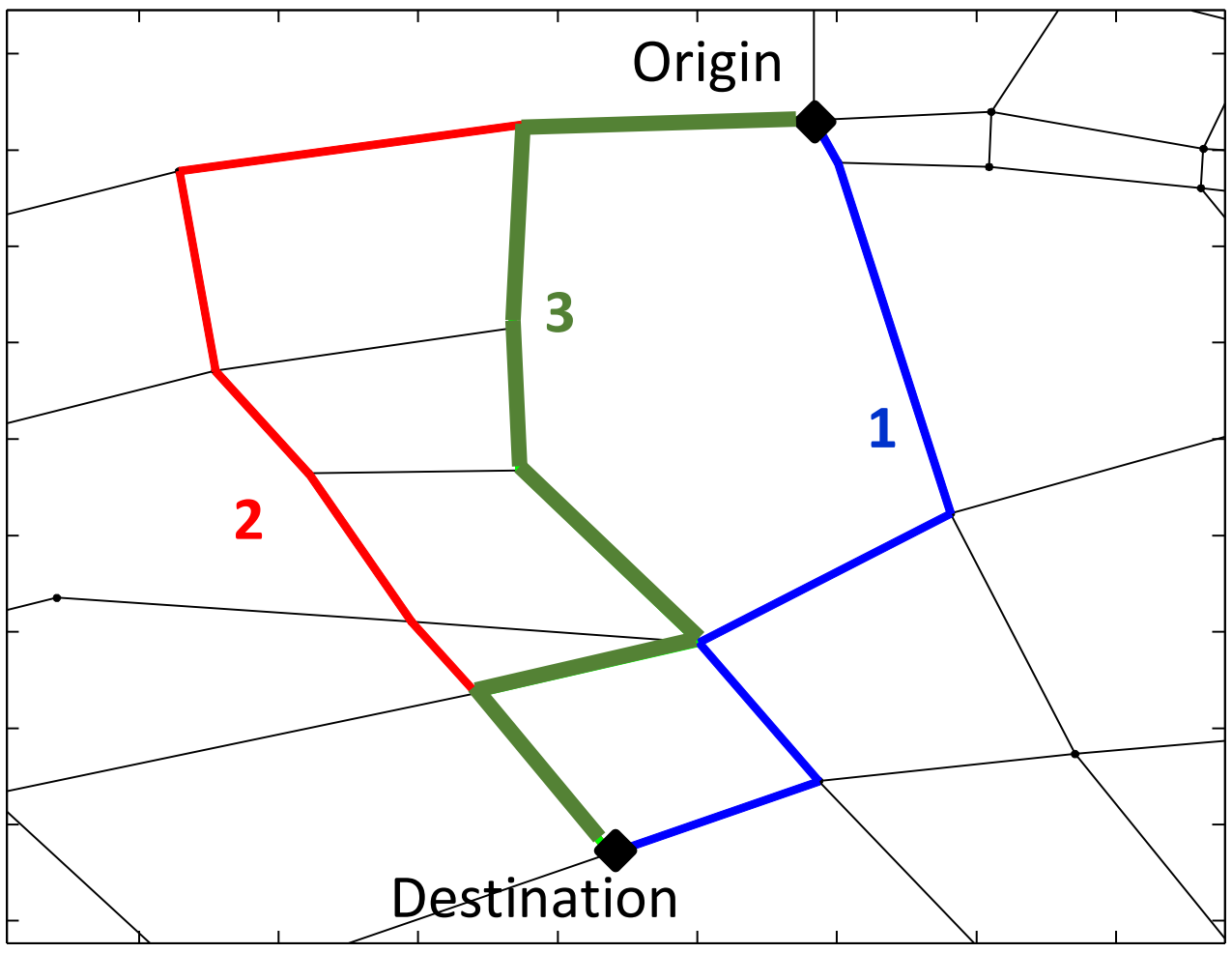

**Figure 6:** VPCRO gives different least cost routes for different vehicle types

This demonstrates that it is important to consider vehicle powertrain types during route optimization. The conversion factor of a vehicle under various traffic conditions is a key factor for cost-efficient routing. The shortest distance routing strategy does not include information about vehicle dynamics and powertrain. It is likely for the shortest path strategy to choose a segment with heavy traffic. If we analyze the paths shown in Figure 6, we can see from Figure 7 that the shortest path goes through a traffic jam and all the segments are under heavy traffic conditions. It can be seen that CV takes the longest path, which mostly consist of segments with low traffic conditions where it performs most efficiently (CV engine has low efficiency in low torque, low speed conditions). On the other hand, HEV is more efficient than CV, so the least cost path for HEV can tolerate more traffic compared to CV but not as much as that of the shortest path in this case.

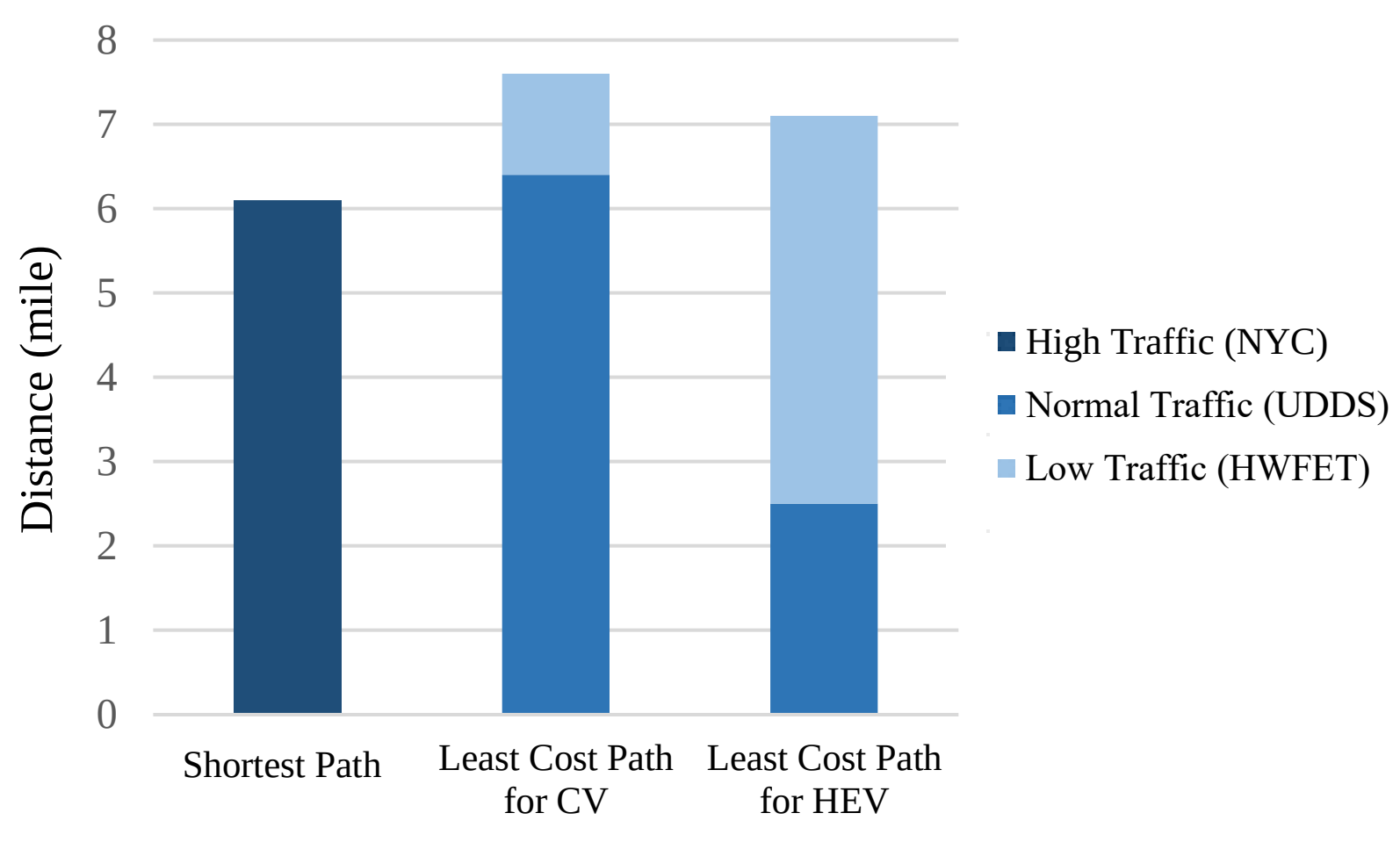

**Figure 7:** Distance and traffic composition of the route for different vehicle types for a sample trip

In Figure 8, we analyze the traffic composition of the average VPCRO given route against that of shortest path and shortest time given. The red, green and yellow segments represents represent heavy (NYC), normal (UDDS) and low traffic (HWFET) conditions, respectively. Since CVs have higher conversion factors in low traffic conditions such as HWFET drive cycle, VPCRO prefers routes that are longer with less traffic compared to shortest path. Electrified vehicles have highest conversion factors under UDDS driving cycle, so for them the majority of the optimal routes consists of traffic conditions represented by UDDS. In Figure 8b, we analyzed the time spent with the same method used in distance comparison and we found out that the total time spent in the optimal path was most for HEVs and the least for CVs as a result of HEVs preferring congested roads which take more time to cover, compared to CVs.

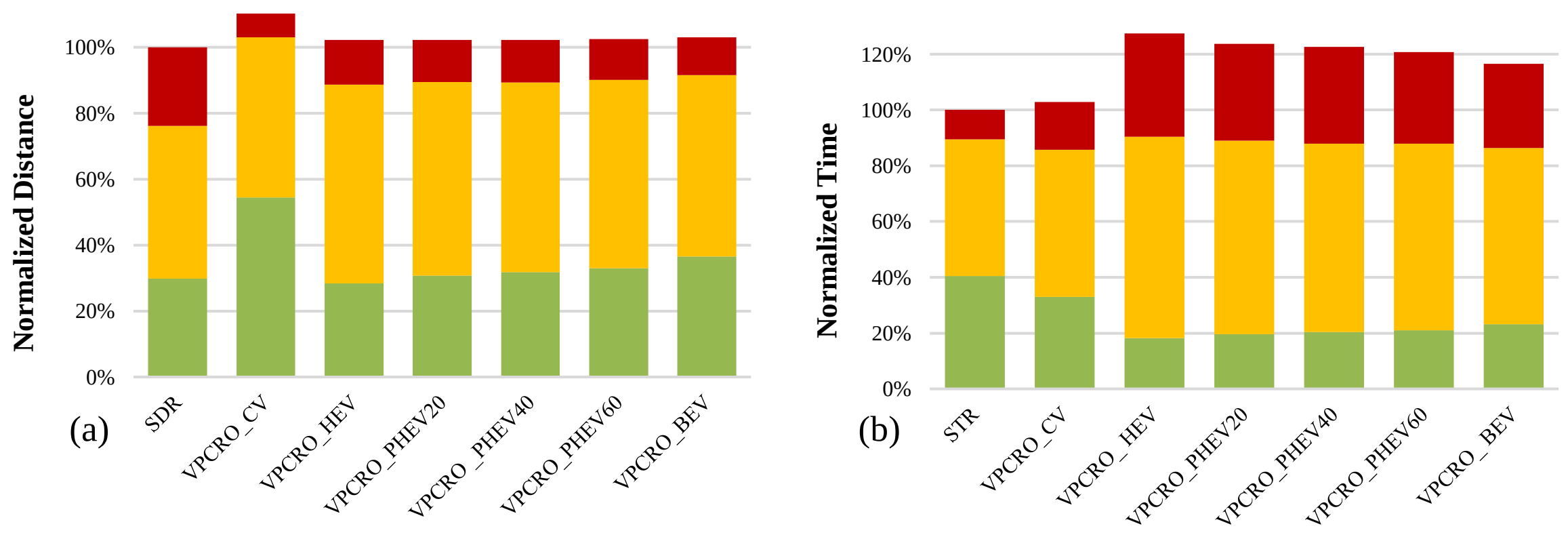

**Figure 8:** (a) Traffic composition of the average VPCRO provided route against that of shortest path. (b) Traffic composition of the average VPCRO provided route against that of shortest time for different vehicle powertrains.

### 4.2. Impact of initial SOC on optimal routing

We find that optimal routes might be significantly different for electrified vehicles with different initial battery state of charge (SOC). It is observed that 3% and 10% of trips have different optimal paths compared to shortest distance method when initial battery SOC changed from 90% to 60% and from 90% to 40%, respectively. Relative cost benefit of VPCRO for EVs with different initial SOCs is higher when VPCRO is compared to shortest time rather than shortest distance. VPCRO becomes even more important for PHEVs with low initial SOC.

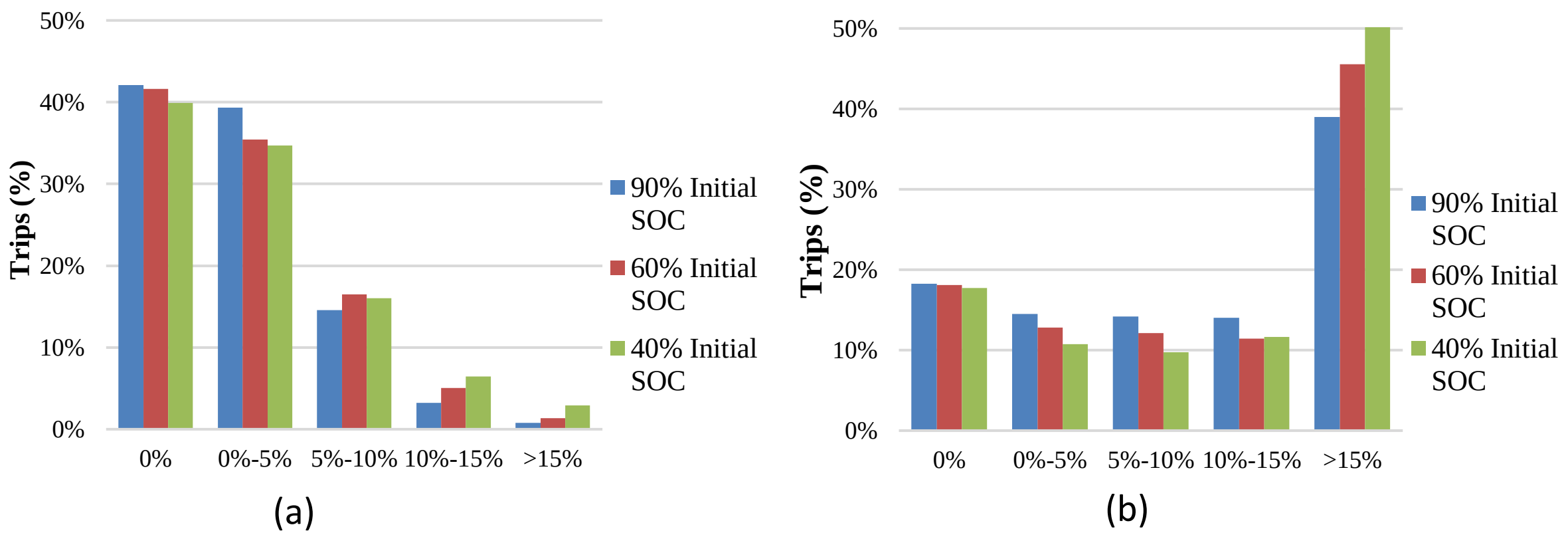

**Figure 9:**(a). Distribution of cost savings via VPCRO for PHEV20 with different initial SOCs compared to Shortest-Distance. (b) Distribution of cost savings via VPCRO for PHEV20 with different initial SOCs compared to Shortest-Time

In Figure 10, for a given origin and destination on the network model, the blue line (1) mark the shortest-path route while the red (2) and green (3) lines mark the least-cost optimal paths when the initial SOC is 90% and 40% in a PHEV20, respectively.

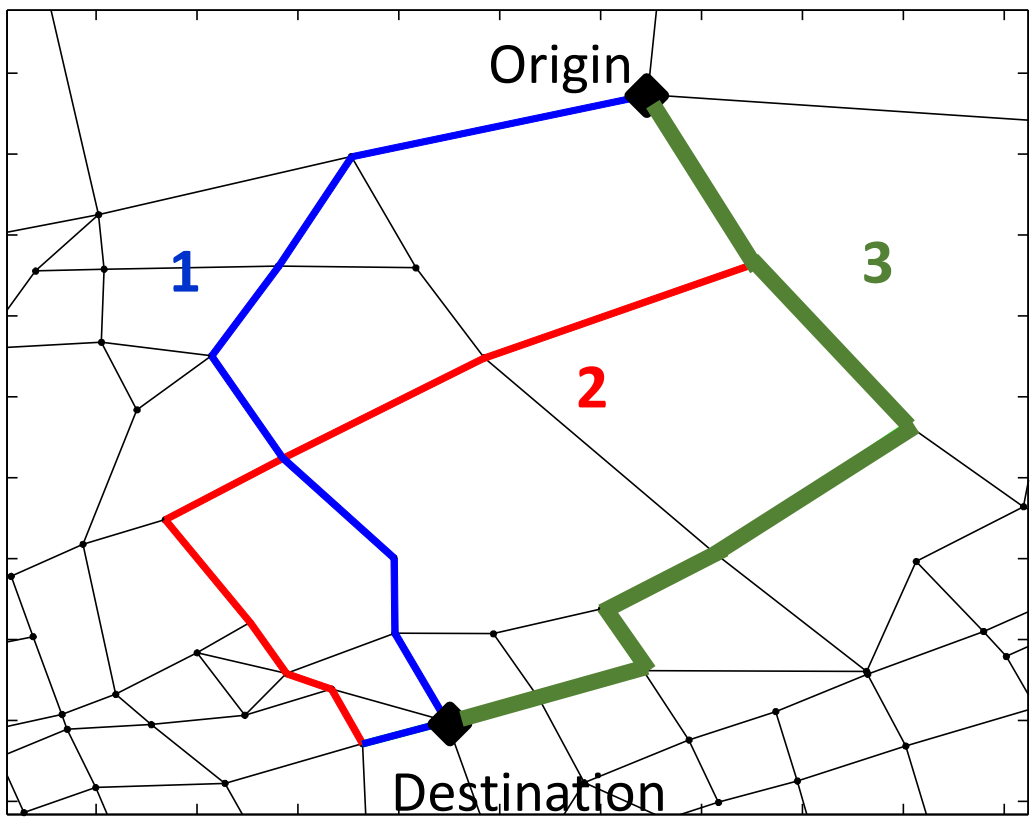

**Figure 10:** Different optimal routes for the same electrified vehicle with different initial battery SOC

In Figure 11, each path has been broken down into segments and the traffic intensity has been compared with the path length. It is seen that with lower initial state of charge, the vehicle depends more on the engine and prefers lower traffic paths, and with higher initial state of charge, the vehicle can use its motor more freely and it can choose average and heavy traffic roads due to the higher efficiency of the electric motor.

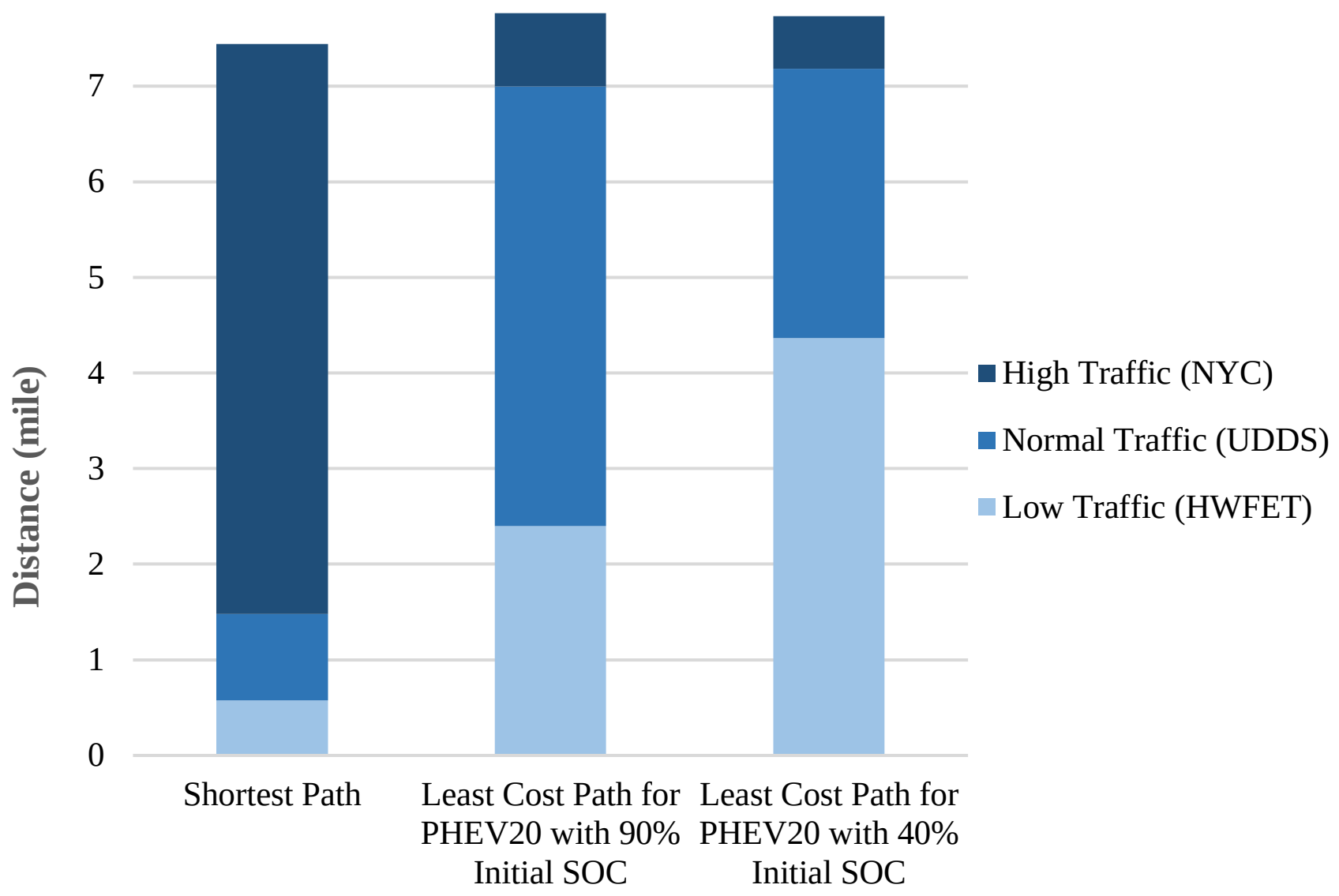

**Figure 11:** Distance and traffic trade-off while routing for an electrified vehicle with different initial battery SOC for a sample trip

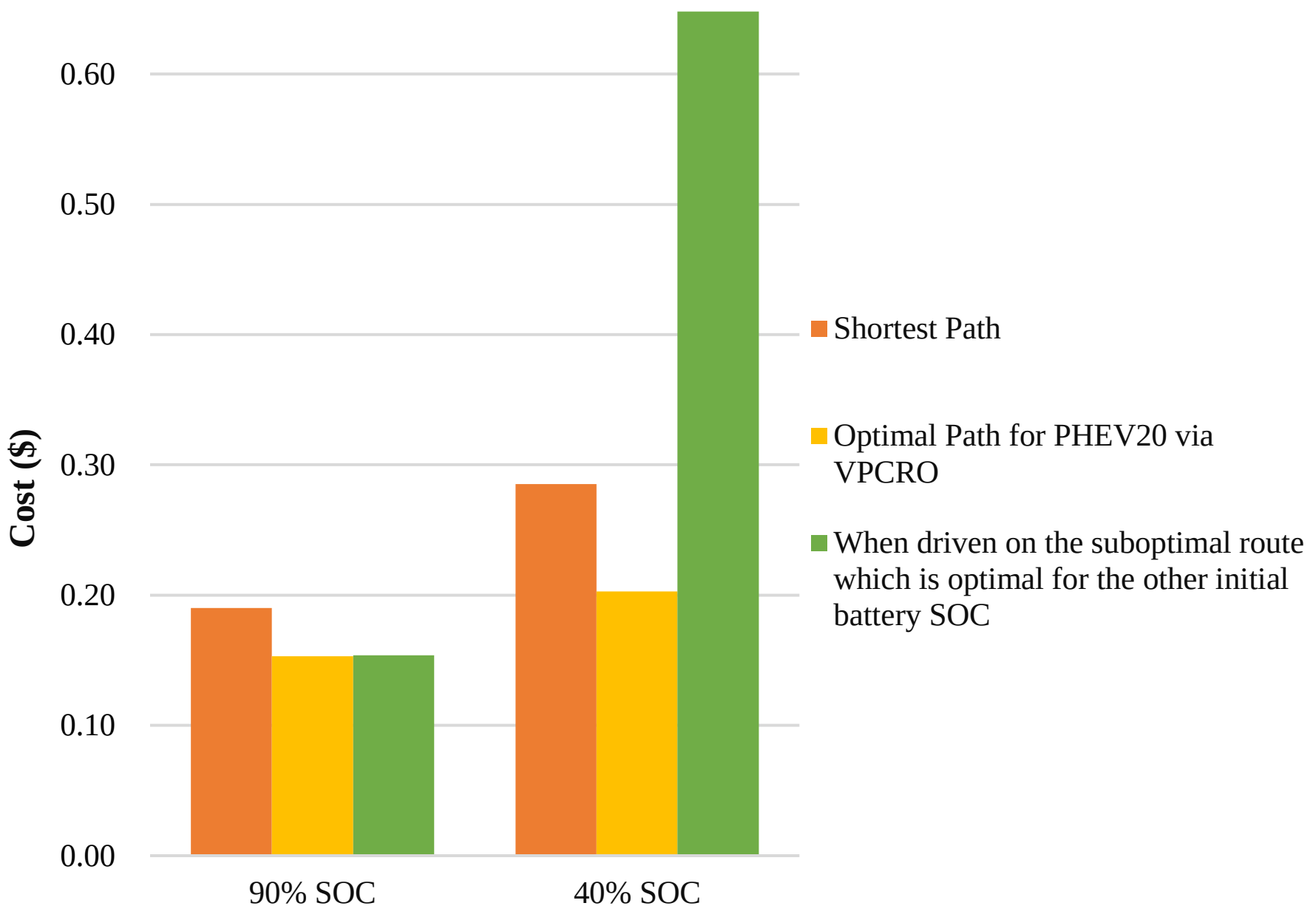

**Figure 12:** Cost comparison of an optimal path for PHEV20 with different initial battery SOC

Figure 12 represents the comparison of cost for the optimal paths for PHEV20 with different initial SOC conditions under shortest-path and the VPCRO cost-optimal path. The figure shows that with the proposed method, the cost incurred gets reduced by 17% and 28% when compared to the shortest path algorithm at 90% and 40% initial SOC conditions, respectively. Similarly, with the proposed method, the total cost gets reduced by 70% for an PHEV20 with 40% initial SOC compared to the cost of traveling the path generated for a PHEV20 with 90% SOC. However, the costs calculated by the proposed method was the same as that incurred by the suboptimal path when the initial SOC was increased to 90%. With higher SOC, the vehicle mostly uses the motor instead of engine, the flexibility and higher efficiency of the motor gives similar costs for both of the routes. However, with lower SOC, the vehicle relies more on the engine and acts more like a HEV, so the average and heavy traffic segments chosen for the 90% SOC results in significantly higher costs for 40% SOC.

### 4.3. Impact of VPCRO on travel time

Figure 13 shows the average travel time and the traffic composition of the optimal routes given by VPCRO routing strategy compared to that of Shortest Distance Routing (SDR) strategy for different vehicle powertrains. Green, yellow and red bars represents the time spent on low-traffic, normal traffic and congested traffic, respectively. It is shown that VPCRO might reduce the travel time for CVs since it prefers paths with less traffic rather than shortest paths with potentially heavy traffic. For hybrid, plug-in hybrid and electrified vehicles, VPCRO might increase the travel time up to 20% compared to SDR. VPCRO benefits CVs more both in terms of reducing travel cost and time compared to electrified vehicles.

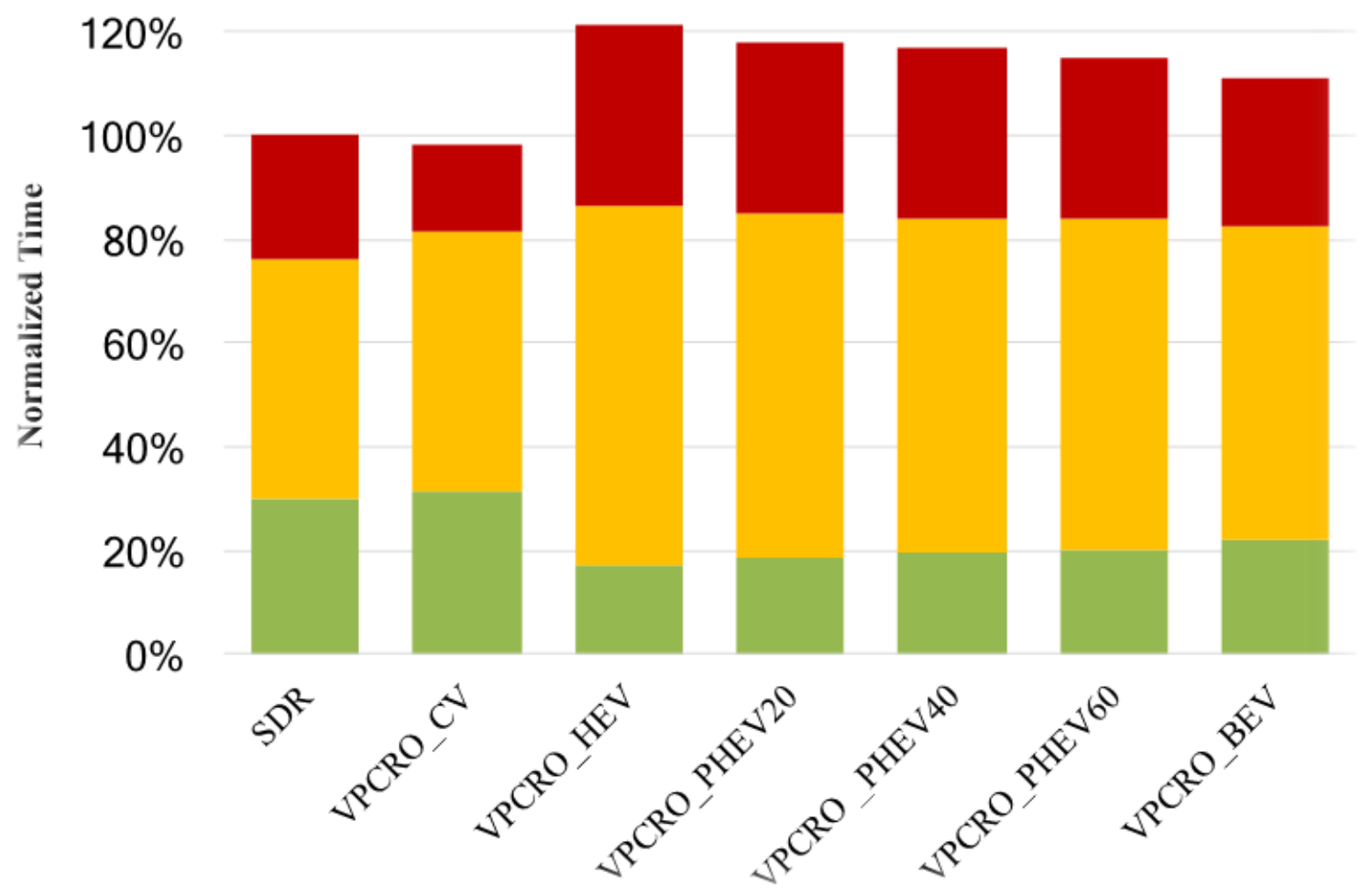

**Figure 13.** Impact of VPCRO on travel time for different vehicle types

## 5. CONCLUSION

In this paper a novel vehicle routing strategy, Vehicle Powertrain Connected Route Optimization (VPCRO), is proposed where vehicle routing decisions benefits from the sensory information from powertrain as well as external information such as traffic conditions. Routing decisions for advanced vehicles are not only influenced by distance but also other factors such as vehicle powertrain type, battery state of charge and the change of component efficiencies under traffic conditions which are not considered by traditional routing systems. The impact of the trade-off between distance and traffic on the cost of the trip might change with the type of vehicle technology and component dynamics. As a result, the least-cost paths might be different from the shortest-distance, least-traffic or least-time paths. We use Dijkstra's Algorithm to implement VPCRO and compare the cost benefits of our algorithm to that of shortest distance and shortest time algorithms for a transportation network model of Shanghai. We find that optimal paths might change significantly for different vehicle types and electrified vehicles with different initial battery state of charge (SOC). For conventional vehicles, up to 80% of routes change with cost savings up to 60% and meanwhile for electrified vehicles, about 60% of routes change with cost savings up to 30%. VPCRO prefers longer paths with less traffic for CVs but shorter paths with higher traffic for HEV due to the high efficiency in traffic conditions of HEV. On the other hand, it is observed that 3% and 10% of trips have different optimal paths, when initial battery SOC changed from 90% to 60% and 40%, respectively. The plug-in electric vehicles with lesser initial battery SOC may benefit more from VPCRO. Specifically, when plug-in electric vehicles do not have high battery SOC, the VPCRO routing system may prefer routes with more UDDS and less HWFET traffic conditions than the case they are fully charged. Given the size of transportation sector, VPCRO routing system can contribute towards significant cost savings for all types of vehicles.

## 6. LIMITATIONS AND FUTURE WORK

Some limitations in this paper might have an impact on the reported benefits of VPCRO approach. This work is mainly focusing on extended-range electric vehicles (EREV) PHEVs instead of blended operation PHEV as the latter varies significantly with control strategy parameters [28][29][30][31]. For electric vehicles, the battery life may affect the routing decision as it influences the efficiency of the vehicle. In this paper, battery is assumed to be in good conditions without degradation. Moreover, as the range of distance in the network selected, it is assumed that BEV can operate under CD mode during the whole trip (no trip is longer than the distance BEV could cover). Traffic conditions in traffic network model are taken from Google maps, however the driving patterns of vehicles are assumed according to this information. Actual, speed-time profiles might be different, however authors do not expect a significant change in the conclusions. Also, we have chosen the traffic situation from Google Map for time period of 8:00 a.m. to 8:30 a.m. to account for major traffic variation, however the reported benefits of VPCRO might change with the level of traffic at different time periods of the days.

Different vehicle powertrain architectures such as series, parallel, and split would have different efficiencies which might impact the routing decisions. Our paper focuses on extended range split powertrain architecture for all hybrid and electrified vehicles. Any factor that changes the efficiency of the vehicle might have an impact on the routing. In this paper, we kept the powertrain architecture constant across all electrified vehicles and investigated the impact of vehicle type, and initial SOC conditions on the routing decisions under different traffic conditions. Some other factors such as powertrain architectures, weather conditions, terrain, and so on are left for future work. At the same time, our proposed framework includes an On Board Diagnostics (OBD) scanner which is capable of getting the real time fuel consumption rate. Real time fuel consumptions happen as a result of all these factors interacting with each other. In this regard, these effects can be accounted for depending on the way this framework is implemented. For offline implementation, efficiency data from different types of vehicles should be stored on the cloud computing system that performs the optimal routing.

Also, further research is needed to determine the additional cost that might be introduced in case many drivers choose the same strategy for route optimization. This might result in dynamically changing congestion of certain segments of the traffic network. Toll cost is shown in the formulation, in this work it has been omitted since toll roads are very specific to certain networks. In real world implementation, this cost item can be easily added to the objective function. BEVs traveling in the Shanghai network model can complete their trips on single charge. Location of charging stations has not been considered in this work.

Although, travel time is investigated here to some extent, real preferences of the drivers might be different and should be assessed. While some might prefer cost minimizing routing, some might prefer the shortest time trips. In future work, we plan on building a system that can manage competing preferences of drivers.

## 7. ACKNOWLEDGEMENTS

We would like to thank to Mr. Vikram Bhattacharjee, a double degree MS student at SYSU-CMU Joint Institute of Engineering, and Mr. Övünç Tüzel, a summer intern from Sabanci University of Turkey, for their input to this work. We are thankful to the members of the Vehicle Electrification Group (VEG) at Carnegie Mellon University and the Laboratory for Intelligent Vehicles and Energy Systems (LIVES) at SYSU-CMU JIE for their valuable feedback.